\documentclass[12pt,a4paper]{iopart}

\usepackage{harvard}
\newcommand{\quant}[2]{$#1\,$#2}
\newcommand{\be}{\begin{equation}}
\newcommand{\ee}{\end{equation}}
\newcommand{\eqref}[1]{(\ref{#1})}

\newcommand{\text}[1]{\mathrm{#1}}

\usepackage{graphicx}

\newcommand{\citet}[1]{\citeasnoun{#1}}
\newcommand{\citep}[1]{\cite{#1}}
\usepackage{textcomp}

\begin{document}

\title[Acceleration of atomic dynamics due to localized energy depositions\dots]{Acceleration of atomic dynamics due to localized energy depositions under X-ray irradiation}

\author{Michael Leitner\footnote{now at: Heinz Maier-Leibnitz Zentrum (MLZ), Technische Universit\"at M\"unchen, Lichtenbergstr.\ 1, 85748 Garching, Germany}, Markus Stana, Manuel Ross and Bogdan Sepiol}
\address{Universit\"at Wien, Fakult\"at f\"ur Physik, Boltzmanngasse 5, 1090 Wien, Austria}
\ead{michael.leitner@frm2.tum.de}

\begin{abstract}
The effect of the absorption of hard X-ray photons on the solid-state dynamics, at energies below the defect creation threshold, is considered. First, it is shown that due to the sensitivity to parameter choices, the present state of the theoretical description of photon-electron-lattice coupling cannot decide on whether the local energy deposition due to the absorption of single photons leads to an acceleration of atomic dynamics at fluences before macroscopic effects due to sample heating set in. Second, a direct study of the dynamics under different incident photon fluxes by atomic-scale X-ray Photon Correlation Spectroscopy (aXPCS) is reported, which provides an upper bound of 0.005 additional atomic jumps per absorbed photon of \quant{8}{keV}. The relevance of such considerations for emergent experimental methods, such as studying atomic dynamics using X-rays as probes, is pointed out. 
\end{abstract}

 
\maketitle

\section{Introduction}
The damaging effects of particle irradiation on matter is a classical field of study and nowadays well understood. This is due to its technological relevance for reactor steels and fusion devices, but also due to its use as a tool for testing atomic-scale concepts in fundamental physics such as, e.g., the migration enthalpy of point defects via post-irradiation annealing recovery \citep{slaterjapplphys1951,kellyirradiationdamage1966}. Damage creation, be it in an accelerator, nuclear reactor or electron microscope, can in general be understood by classical concepts on a local scale, in that a single primary particle transfers part of its kinetic energy onto an atom in a two-body collision. If this energy exceeds some well-defined threshold on the order of \quant{25}{eV} in solid matter \citep{kenikphilmag1975}, the atom will leave its site to create a Frenkel pair, or the ensuing collision cascade will even melt the crystal locally. 

In contrast to such microscopic effects, the typically encountered perturbations induced in the sample by electro-magnetic irradiation of all but the highest gamma energies belong to the macroscopic domain. Apart from issues such as single-shot sample vaporization by XUV radiation \citep{chapmannatphys2006}, the transient modifications to the electronic system observed in, e.g., fast electronic relaxations studied by pump-probe techniques \citep{bonnprb2000,pfaunatcommun2012} or laser-induced athermal \citep{tomprl1988,medvedevprb2013} and thermal \citep{hauriegeprl2012} phase transitions are fields of active study.

In scattering experiments, where the radiation beam is used as a probe, it is imperative to know about the beam's effects on the sample. The fact of radiation damage in electron microscopy is well-established \citep{urbanphysstatsola1979}, and also hard X-ray irradiation can affect soft matter, specifically biological systems \citep{kirzqrevbiophys1995}. On the other hand, there is a silent consensus that hard matter (such as metals, semiconductors and ceramics) is unperturbed by the absorption of single X-rays, so that the possibility of an influence of the beam on the sample in this case is only rarely mentioned, let alone investigated. Note that the sample modification under pulses from free-electron lasers \citep{hauriegeprl2012,hruszkewyczprl2012} is explained by the high energy deposition densities and quite expected.

The observed absence of beam damage (permanent modifications of the sample's macrostate) under low-fluence X-ray irradiation is no sufficient condition for the sample \emph{dynamics} to be unaffected, however. While in the case of biological systems the photons enable the sample to lower its free energy by breaking bonds and thereby accumulate irradiation damage, in crystalline matter the principal radiation-induced defect is a Frenkel pair and as such generally thermodynamically unstable. Therefore the following scenario is a priori perfectly possible: the above-mentioned and often-quoted knock-on threshold of about \quant{25}{eV} only decides whether long-range atomic displacements and thereby long-lived Frenkel pairs result, but crystalline order can be broken locally also by the absorption of photons of somewhat lower energy. Once this energy had dissipated, the order would be re-established, thus resulting in a few net atomic exchanges. Further, at still lower energies a local enhancement of thermal diffusion due to the dissipation of an X-ray photon's energy into the lattice is conceivable, which could have appreciable effects on the overall diffusion rate due to the very non-linear dependence of the Boltzmann factor on temperature.

These issues can be put succinctly in the following question: does the absorption of hard X-ray photons at dose rates where sample heating can be neglected lead to a noticeable increase in the atomic jump rate? As will be elaborated below, available theoretical treatments are not able to give an answer. Also, in conventional tracer measurements, the sensitivity of the method is a limiting factor, and consequently no positive results have been reported. However, the newly developed method of atomic-scale X-ray Photon Correlation Spectroscopy \citeaffixed{leitnernmat2009}{aXPCS,} affords the possibility to test the situation with unprecedented resolution. This technique studies diffusive dynamics at atomic length-scales by monitoring the temporal evolution of the intensity pattern of scattered coherent radiation  and is one of the proposed fields of study for the new generation of hard X-ray free electron lasers such as the LCLS at Stanford \citep{robertjphysconfser2013} and the European XFEL at Hamburg \citep{gruebelnimb2007}. Apart from the implications on electron-lattice coupling on the local scale, the study presented here is therefore also of high relevance for the interpretation of experimental determinations of dynamics with X-rays as probes.

In the classical literature on particle and gamma irradiation-induced diffusion enhancements, one distinguishes between the concepts of ion beam mixing, which concerns the mobility of atoms in the collision cascade immediately following the interaction of a primary particle with the lattice \citep{averbacknimb1986}, and radiation-enhanced diffusion, corresponding to accelerated thermal diffusion thanks to the increased density of point defects during and after irradiation \citep{dugdalephilmag1956,dienesjapplphys1958}. Direct observations of enhanced diffusion by tracer measurements are scarce, and require large doses for measurable effects \citep{leephilmaglett1994}.

Here we will define three scenarios for the effect of X-ray irradiation on a sample's atomic dynamics, only two of which are conventionally considered. The third scenario, where the energy deposited locally into the lattice leads to a spatially and temporally confined acceleration of dynamics, will be treated here for the first time. We will elaborate that the existing theoretical and experimental data lead to uncertainties of the proposed effects on orders of magnitude, and cannot be used to rule them out. Specifically we will use a typical aXPCS experiment at a third-generation synchrotron for giving quantitative estimates. Also, we will report on a dedicated experiment on these issues, with a sensitivity down to well below one atomic jump per absorbed photon.

\section{Definition of scenarios and estimates of their effects}
For the scenarios to be considered here, we will use parameter values that correspond to those of a typical experiment as reported below.

\subsection{Macroscopic heating}
The most basic effect of relevance is the macroscopic heating of the sample by the beam, which has already been considered in the context of experiments at free electron lasers by \citet{gruebelnimb2007}. We assume the sample to consist of a \quant{10}{\textmu m} thin metal foil in a beam of about $2\cdot 10^{11}$ photons of \quant{8}{keV} per second, focused to a diameter of \quant{5}{\textmu m}. For an experiment in transmission geometry, about half of the photons will be absorbed in the sample, corresponding to a heating power of the beam of \quant{0.13}{mW}. Neglecting heat radiation, using the thermal conductivity of Cu at \quant{380}{W/(m\,K)} and solving the two-dimensional heat conduction equation with circular symmetry gives a temperature difference between the edge of the illuminated area at \quant{r_0=2.5}{\textmu m} and the sample holder at \quant{r_1=2}{mm} of \quant{\Delta T=0.036}{K}. Assuming a homogeneous illumination within the focal spot gives an additional temperature difference of \quant{0.003}{K} between the beam centre and $r_0$. The measured activation energy of \quant{2.1}{eV} \citep{leitnernmat2009} corresponds at the relevant temperature range roughly to a doubling of the jump frequency for each \quant{9}{K}. A heating of \quant{0.04}{K} due to the beam can therefore safely be neglected.

Additional to the steady-state heating, the bunched temporal structure of a synchrotron beam and the resulting temperature spikes have to be considered. We take as an extreme scenario a time structure of \quant{200}{ns} between bunches at full integrated current, such as used for studying nuclear transitions in the time domain \citep{sepiolprl1996}. A typical number density of atoms in a metal of \quant{5\cdot 10^{28}}{m$^{-3}$} gives $10^{13}$ atoms within the illuminated volume and therefore a Dulong-Petit heat capacity of \quant{4.1\cdot 10^{-10}}{J/K}. Thus, the adiabatic sample heating during the absorption of a photon bunch corresponds to temperature spikes of \quant{0.063}{K}. Again, such a temperature increase can be neglected, especially given the fact that the synchrotron beam time structure used for XPCS measurements typically corresponds to much higher bunch repetition rates and therefore smaller spikes. On the other hand, the high single-pulse fluences of X-ray free electron laser sources has indeed been observed to lead to appreciable effects \citep{hruszkewyczprl2012}.

\subsection{Knock-on displacements}
Having excluded macroscopic effects, we now turn to the microscopic situation. While by the Compton formula \citep{comptonpr1923} an \quant{8}{keV} photon can transfer at most \quant{250}{eV} to a free electron, the equivalent figure for atoms is on the order of meV. This shows that, contrary to all previous studies of radiation damage by particle and MeV gamma quanta \citep{dugdalephilmag1956}, this scenario can be ruled out for X-rays.

\subsection{Microscopic dynamics enhancement}
The third scenario concerns a local acceleration of diffusive dynamics due to the transient microscopic heating after the absorption of an X-ray photon. This is conceivable, as the Boltzmann factor governing diffusion is highly non-linear, so that the mean atomic jump rate is not simple given by the mean sample temperature (which is only very weakly affected by the beam, as shown above), but can be sizeably increased by the local and transient heat spikes. The relevant space- and time-scales and the expected effects will be discussed below.

The pathway of photon energy dissipation into the lattice is as follows \citep{medvedevprb2013}: In a first stage the photon loses its energy to excitations of the electronic structure of the solid. This happens via the ejection of a photoelectron, perhaps after a few events of Compton scattering. The excited atomic state can decay via again emitting a fluorescence photon, thereby restarting this process with somewhat lower energy, or emitting an Auger electron. This happens on electronic timescales, i.e., femtoseconds, after which the energy has been conferred to a few hot electrons in the keV range as determined by the core levels of the constituent elements.

The interactions of hot electrons with the unperturbed lattice and electronic system in a later stage can be classified into elastic collisions with the ionic cores and inelastic interactions with the electrons. The latter include the emission of low-energy secondary electrons and plasmons, with a corresponding inelastic mean free path that decreases from around \quant{30}{\AA} at \quant{2}{keV} to around \quant{5}{\AA} at \quant{50}{eV} \cite{echeniquechemphys2000,tanumasurfinterfanal2011}. At lower energies quantitative theoretical predictions become difficult, but the observed increase in escape depths \citep{seahsurfinterfanal1979} agrees with the picture of delocalized quasi-particles with long lifetimes at low energies. 

Tabulated free-atom calculations of the elastic scattering, e.g.\ by \citet{czyzewskijapplphys1990}, show that the elastic mean free path is in general shorter than the inelastic mean free path (below \quant{10}{\AA}), but corresponds for high electron energies mostly to small scattering angles and therefore near-ballistic transport. However, around \quant{100}{eV} significant scattering angles become probable, so that electrons in this energy range will figuratively bounce from atom to atom. At even lower energies the mean free path would decrease below atomic distances. This logical problem is relieved by the fact that the de Broglie-wavelength then becomes appreciable, so that delocalized states, which do not scatter off a perfect lattice, become the appropriate description, in accordance with above-referenced increasing escape depths.

The critical issue is the spatial scale of above-mentioned processes: While both the initial keV-range electrons are mobile enough so that overlaps of their effects seem improbable and also the electron-phonon interaction at near-thermal energies is delocalized, electrons in the range of \quant{100}{eV} will undergo repeated elastic collisions, each with an energy transfer of a few meV, within a small spatial region \citep{dingscanning1996}. It is hard to quantify the corresponding temperature increase due to the lack of pertinent theoretical results. A very rough estimate of a single \quant{100}{eV} electron losing half of its energy to ionic motion within a region of \quant{25}{\AA} diameter would give a temperature increase of \quant{470}{K} or \quant{0.12}{eV/atom} for about 410 atoms. To our knowledge, detailed simulations of the spatio-temporal behaviour of the electronic excitation have been done only in the insulator diamond, giving dimensions of the electron cloud on the order of \quant{100}{\AA} after \quant{90}{fs}, when the mean electronic energy has decayed to a few eV \citep{ziajajapplyphs2005}. In comparison, we would expect the spatial scales in a typical elemental metal to be pronouncedly smaller, or even more so in the case of a disordered solid solution.

The timescale of the dissipation of the electron energy into the lattice is on the order of a few phonon cycles \citep{elsayed-aliprl1987}. On the other hand, this is clearly also the timescale that applies to the lattice cooling of a microscopic hot region, as additional contributions from phononic heat conduction are limited by the speed of sound. The above-mentioned localized and transient temperature increase of \quant{470}{K} corresponds to an enhancement of the local dynamics by a factor of about $8\cdot 10^8$ at a macroscopic temperature of \quant{550}{K} and the above-quoted activation energy of \quant{2.1}{eV}. The experimental parameters of $10^{11}$ absorbed photons of \quant{8}{keV} per second, each corresponding to 80 secondary electrons of \quant{100}{eV}, over an illuminated volume of $10^{13}$ atoms leads to a given atom experiencing such an enhancement about 330 times per second, or for a fraction of $10^{-9}$ of the time. As a consequence, these assumptions would give an overall acceleration of the dynamics in the sample by a factor of about two.

However, due to the non-linearity of the Boltzmann factor as a function of temperature the uncertainty of these estimated figures is very large: a deviation on the order of 10\% in the spatial scale of the energy dissipation gives a correspondent deviation in the temperature rise, which leads to an order-of-magnitude difference in the diffusion enhancement. Also local inhomogeneities in the heating due to either the random walk of the secondary electrons within a dissipation region or overlapping dissipation regions give rise to further uncertainties. 

Concluding the theoretical part of this work, we have demonstrated by simple arguments that the energy brought into the system by the beam on the macroscopic scale is not sufficient for dynamics to be noticeably accelerated. However, we cannot decide on the influence of microscopic effects, which is due to the lack of detailed theories of the thermal excitation of the solid after the absorption of an X-ray photon, resolved in space and time. We therefore turn to direct experimental investigations of the issue at hand.

\section{Experimental results}
We performed an atomic-scale X-ray Photon Correlation Spectroscopy experiment with a sample of single-crystalline Cu$_{90}$Au$_{10}$ dedicated to probe the effect of incident X-ray irradiation on the observed dynamics. This alloy has already been the subject of a previous publication \citep{leitnernmat2009}, where details about the sample preparation, experimental set-up, and data evaluation are given. Here, however, we employed the superior available coherent X-ray intensity at the beamline P10 at PETRA III, which allowed us to vary the incident intensity while still being able to measure the dynamics with a high signal-to-noise ratio. 
\begin{figure}
\centerline{\includegraphics{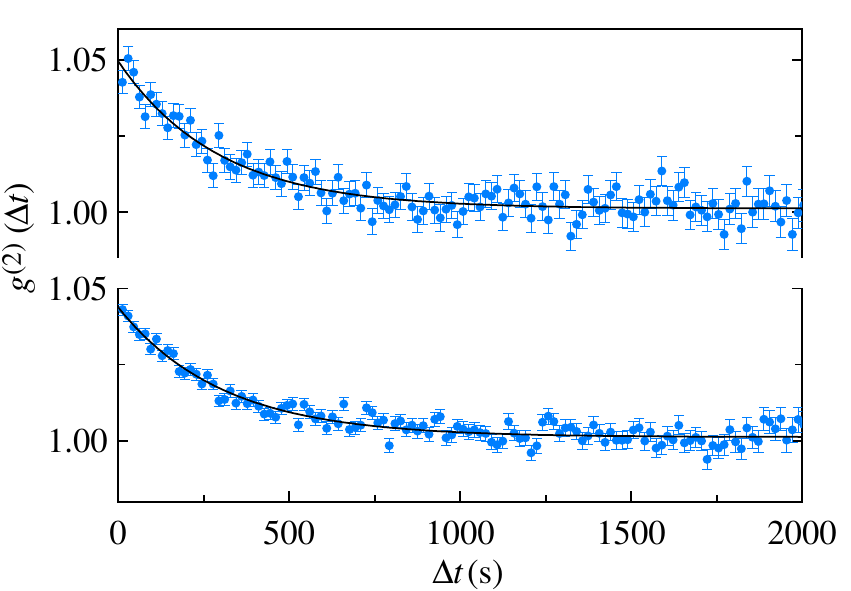}}
\caption{Auto-correlation function at \quant{553}{K} and $2\theta=5^\circ$ for full beam (lower plot) and \quant{100}{\textmu m} Si attenuation (upper plot), together with exponential fits.}\label{acf}
\end{figure}

At \quant{8}{keV}, the intensity of the full beam corresponds to the above-quoted values of $2\cdot 10^{11}$ photons per second, focused to a diameter of \quant{5}{\textmu m}. The beam can be attenuated by absorbers of Si, which has an attenuation length of \quant{69.6}{\textmu m} at \quant{8}{keV}. The experiment was done with the sample mounted in two different furnaces at nominal sample temperatures of \quant{543}{K} and \quant{553}{K}, respectively. Instead of doing single long continuous measurements, we performed many short measurement runs and changed between attenuations in a random way. This allows us to rule out a compromising of the measured dynamics by temporal instabilities such as sample relaxations, as the obtained decays for a given attenuation varied only within their estimated errors. For further evaluation the auto-correlation functions corresponding to a given attenuation were averaged with the appropriate weights. In order to cover a range of correlation times, we also varied the scattering angle $2\theta$ between $5^\circ$ and $15^\circ$.

Auto-correlation functions for full and attenuated beams are given in Fig.~\ref{acf}. The high signal-to-noise ratio is evident, and, as expected for a crystalline system, the data can be described by a simple exponential decay according to
\be
g^{(2)}(\Delta t)=1+\beta\exp(-2\Gamma\Delta t),
\ee
where $\Gamma$ is the decay constant, dependent on jump mechanism, position in reciprocal space, and temperature, and $\beta$ is the coherence factor that quantifies the degree to which interference due to coherent scattering is visible and that depends on the instrumental set-up.

\begin{figure}
\centerline{\includegraphics{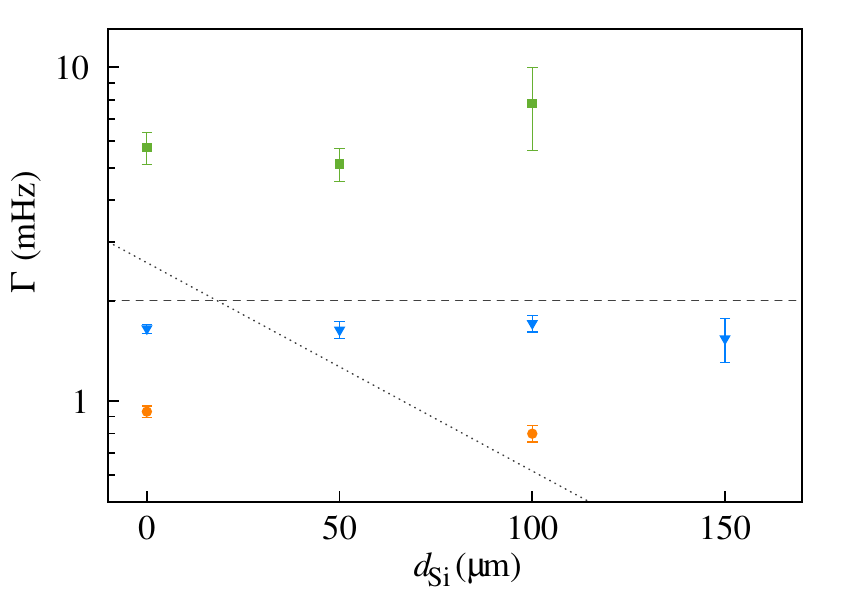}}
\caption{Decay constants $\Gamma$ as a function of Si-attenuator thickness at \quant{553}{K} for $2\theta=5^\circ$ (blue triangles) and $2\theta=15^\circ$ (green squares), and at \quant{543}{K} for $2\theta=15^\circ$ (orange circles) on logarithmic scale, compared to models of purely thermal (dashed line) or irradiation driven (dotted line) dynamics.}\label{gammas}
\end{figure}

The resulting fitted decay constants for a number of attenuator thicknesses are given in Fig.~\ref{gammas}. The highest attenuation corresponds to a factor of about 8.6; it is therefore immediately visible that the effect of a beam-driven acceleration of the dynamics is small, if present at all. 

For a more quantitative statement we make the natural assumption that the enhancement of the atomic jump frequency is proportional to the number of absorbed photons. Thus we model the dependence of the atomic jump frequency $\nu$ on incident flux $I$ and temperature $T$ as 
\be
\nu(I,T)=\nu_\text{th}(T)+\nu_\text{b}I,\label{nuit}
\ee
where $\nu_\text{th}$ is the thermal contribution and $\nu_\text{b}$ is the enhancement due to the beam, most intuitively measured in additional atomic jumps per absorbed photon. The decay constant $\Gamma$ is proportional to the atomic jump frequency
\be
\Gamma(\vec{q},I,T)=a(\vec{q})\nu(I,T),\label{gammaqit}
\ee
where $a(\vec{q})$ depends on the position in reciprocal space $\vec{q}$ and is characteristic of the atomic jump mechanism \citep{leitnerjphyscondmat2011}. Specifically, for the scattering angles relevant here we have $a_{5^\circ}=0.24$ and $a_{15^\circ}=1.23$ in the chosen sample orientation.

\begin{figure}
\centerline{\includegraphics{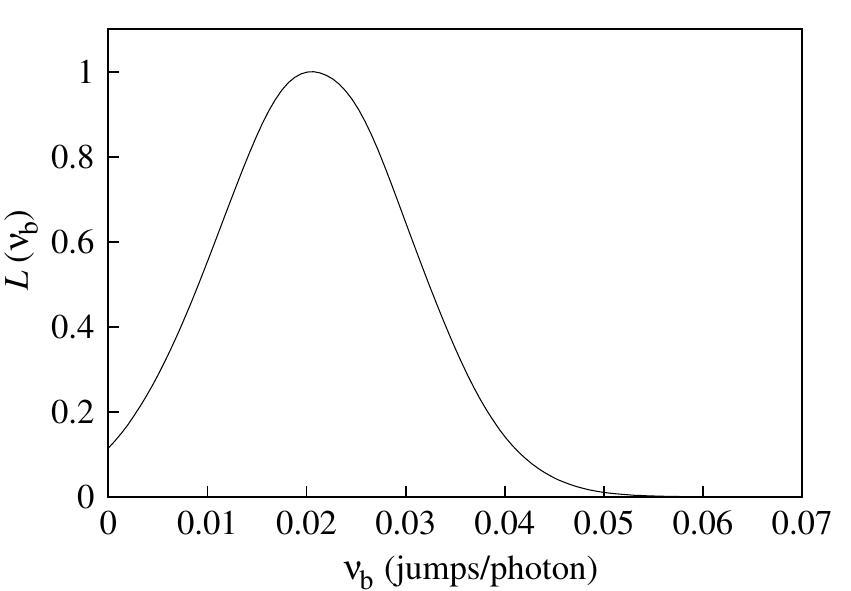}}
\caption{Likelihood function of dynamics enhancement $\nu_\text{b}$.}\label{enhlikelihood}
\end{figure}

Fitting all auto-correlation functions directly and simultaneously with decay constants $\Gamma(q,I,T)$ as specified by the model \eqref{nuit} and \eqref{gammaqit} gives a behaviour of the likelihood function for $\nu_\text{b}$, marginalized over all other free parameters, as illustrated in Fig.\ \ref{enhlikelihood}. According to a maximum-likelihood approach, one would therefore estimate $\nu_\text{b}$ to be positive, although not strongly significant. Its maximum-likelihood value of $\nu_\text{b}^*=0.021$ jumps per absorbed photon corresponds to a full-beam enhancement of 2\% over the thermal dynamics at \quant{553}{K}. 

Because of the low level of significance of this maximum-likelihood estimate, adopting the Bayesian point of view leads to a more useful result: Modelling the prior probability distribution of the local temperature rise as a uniform distribution over some range corresponds via the exponential nature of the Boltzmann factor approximately to a log-uniform prior distribution of the beam enhancement $\nu_\text{b}$, i.e., a distribution according to which the order of magnitude of $\nu_\text{b}$ has uniform prior probability. A sensible lower bound would be the assumption that any given atom suffers at most a single elastic collision with a secondary electron, corresponding to typical energy transfers of \quant{2}{meV} and an overall diffusion enhancement six orders of magnitude below the value of maximum likelihood $\nu_\text{b}^*$. Due to the super-exponential decay of the likelihood function $L(\nu_\text{b})$ above $\nu_\text{b}^*$, the choice of the upper boundary is irrelevant. Accordingly, the experimentally observable values of $\nu_\text{b}$ larger than 0.005 jumps per absorbed photon make up only 11\% of the resulting posterior distribution, so that this value can serve as an upper bound for the effect. This reasoning implies that the observed maximum in $L$ is likely just due to stochastic effects and should not be construed as an indication of experimentally-resolvable beam-driven dynamics acceleration at these fluences.

\section{Conclusion}
We have identified a scenario of an enhancement of diffusive atomic dynamics due to absorption of X-ray photons, and we have shown that current knowledge on the microscopic spatial and temporal features of electron-lattice energy dissipation is insufficient for estimating its relevance. Even though our direct aXPCS measurement is compatible with a weak enhancement, Bayesian inference with an uninformative prior implies that in the present case the effect is probably below the experimental resolution and provides an upper bound of 0.005 additional atomic jumps per absorbed photon of \quant{8}{keV}. The dependence of correlation time on temperature as confirmed in recent aXPCS experiments \citep{leitnerprb2012,stanajphyscondmat2013,rossnewjphys2014} is in accordance with this conclusion. However, for future investigations the possibility of beam-driven dynamics should not be overlooked, as for instance higher energy transfers in elastic collisions together with lower activation energies of diffusion in systems such as aluminium could render it effective. This precaution applies even more so for the increased fluxes at X-ray free electron lasers, but there apparently already the macroscopic transient sample heating due to the single pulse fluence is the limiting factor \citep{hruszkewyczprl2012}. Apart from these implications for aXPCS experiments, it would be very welcome to have further theoretical work on the dissipation of an X-ray photon's energy into the lattice in order to be able to set narrower bounds on the magnitude of the effects discussed here.

\ack
We thank Gero Vogl for providing the initial background on classical radiation damage physics and Michael Sprung, the local contact at PETRA III, for assisting with the measurements. This work was supported by the Austrian Science Fund (FWF): P22402.

\section*{References}

\end{document}